\documentclass[%
 reprint,
 amsmath,amssymb,amsthm,
 nofootinbib,
 aps,
]{revtex4-1}

\usepackage{dcolumn}
\usepackage{bm}
\usepackage[mathlines]{lineno}
\usepackage{graphicx}
\usepackage{csquotes}
\usepackage[final]{changes}

\begin{document}
\preprint{APS/123-QED}

\title{Quantum Scalar Field Theory Based on \replaced{the Extended Least Action Principle} {Principle of Least Observability}}

\author{Jianhao M. Yang}
\email[]{jianhao.yang@alumni.utoronto.ca}
\affiliation{Qualcomm, San Diego, CA 92121, USA}

\date{\today}		

\begin{abstract}
Recently it is shown that the non-relativistic quantum formulations can be derived from \replaced{an extended least action principle}{a principle of least observability}\cite{Yang2023}. In this paper, we apply the principle to massive scalar fields, and derive the Schr\"{o}dinger equation of the wave functional for the scalar fields. The principle extends the least action principle in classical field theory by factoring in two assumptions. First, the Planck constant defines the minimal amount of action a field needs to exhibit in order to be observable. Second, there are constant random field fluctuations. A novel method is introduced to define the information metrics to measure additional observable information due to the field fluctuations, \added{which is then converted to the additional action through the first assumption.} Applying the variation principle to minimize the total actions allows us to elegantly derive the transition probability of field fluctuations, the uncertainty relation, and the Schr\"{o}dinger equation of the wave functional. Furthermore, by defining the information metrics for field fluctuations using general definitions of relative entropy, we obtain a generalized Schr\"{o}dinger equation of the wave functional that depends on the order of relative entropy. Our results demonstrate that the extended least action principle can be applied to derive both non-relativistic quantum mechanics and relativistic quantum scalar field theory. We expect it can be further used to obtain quantum theory for non-scalar fields.
\end{abstract}

\maketitle
\section{Introduction}

Advancements of quantum information and quantum computing~\cite{Nielsen,Hayashi15} in recent decades have inspired active researches for new foundational principles for quantum mechanics from the information perspective~\cite{Rovelli:1995fv, zeilinger1999foundational, Brukner:ys, Brukner:1999qf, Brukner:2002kx, Fuchs2002, brukner2009information, Brukner:vn, spekkens2007evidence, Spekkens:2014fk, Paterek:2010fk, gornitz2003introduction, lyre1995quantum, Hardy:2001jk, Dakic:2009bh, masanes2011derivation, Mueller:2012ai, Masanes:2012uq, chiribella2011informational, Mueller:2012pc, Hardy:2013fk, kochen2013reconstruction, 2008arXiv0805.2770G, Hall2013, Hoehn:2014uua, Hoehn:2015, Stuckey, Mehrafarin2005, Caticha2011, Caticha2019, Frieden, Reginatto, Yang2021}. Reformulating quantum mechanics based on information principles can bring in new conceptual insights to the unresolved challenges in the current quantum theory. For instance, is probability amplitude, or wavefunction, just a mathematical tool or associated with ontic physical property? Does quantum entanglement imply non-local causal connection among entangled objects? With this motivation, recently an extended least action principle is proposed to derive the formalism of non-relativistic quantum mechanics~\cite{Yang2023}. \deleted{ Here observability refers to how observable that a physical object exhibits in its dynamics. It measures the information available for potential observation.} The principle can be understood as extending the least action principle in classical mechanics by minimizing proper information measures. This is achieved by factoring two assumptions. First, there is a lower limit to the amount of action a physical system needs to exhibit in order to be observable. Such a discrete action unit is defined by the Planck constant. It serves as a basic unit to measure the observable information from the action a physical system exhibits during its motion. Second, there is vacuum fluctuation that is completely random. New information metrics are introduced to measure the additional distinguishability, or observable information, due to these random fluctuations, \added{which is then converted to additional amount of action due to the vacuum fluctuations using the first assumption.} Applying the variational principle to minimize the total actions allows us to elegantly recover the basic formulations of non-relativistic quantum mechanics. In addition, a family of generalized Schr\"{o}dinger equation for the wave functional is obtained by defining the information metrics for vacuum fluctuations using generic relative entropy definitions.   

The goal of this paper is to apply the same principle to relativistic quantum field theory. Specifically, we will apply the extended least action principle to derive the quantum field theory of massive scalar fields. Impressively, we find that the only adjustment needed to the extended least action principle is to replace the assumption of random vacuum fluctuations in the non-relativistic setting to random field fluctuations in the relativistic settings. By recursively applying the extended least action principle, we are able to derive the transition probability density of the field fluctuations, the uncertainty relation, and most importantly, the Schr\"{o}dinger equation of the wave functional for the scalar fields. The Schr\"{o}dinger equation of the wave functional is the fundamental equation for the quantum scalar field theory in the Schr\"{o}dinger picture, and it is typically introduced as a postulate. Here we derive it from a first principle. Similarly to the non-relativistic quantum formalism, by relaxing the definition of the information metrics using generic relative entropy, we obtain a family of generalized Schr\"{o}dinger equations. The application of such generalized Schr\"{o}dinger equations needs further investigation, but the result shows the flexibility of the mathematical framework.

The Schr\"{o}dinger picture offers several advantages compared to the standard Fock space description of scalar fields~\cite{Long}. In particular, the Schr\"{o}dinger wave functional gives an intrinsic description of the vacuum without reference to the spectrum of excited states, which is an inherent problem in the Fock space of state in curved spacetime~\cite{Long}. It is also argued that the Schr\"{o}dinger picture in field theory is the most natural representation from the viewpoint of canonical quantum gravity where the spacetime is usually decomposed into a spatial manifold evolving in time~\cite{Corichi}. The Schr\"{o}dinger formulations in both non-relativistic quantum theory and relativistic quantum field theory allows us to understand the difference and similarity between the two theories. It may provide hints on applying certain concepts from one theory to the other. For instance, calculating information metrics such as the entanglement entropy of a quantum field is challenged~\cite{Takayanagi}. In non-relativistic quantum mechanics, such a quantity for entangled systems is typically calculated with the help of the wave function. With the availability of the Schr\"{o}dinger wave functional, one may find a similar method to calculate the entangled entropy for a scalar field.

Extending the least action principle in classical mechanics to derive quantum theories not only shows clearly how classical mechanics becomes quantum mechanics, but also offers a powerful mathematical framework. As shown in this paper, the principle and mathematical framework allow us to derive the Schr\"{o}dinger equation for the wave functional of the scalar field in a way very similar to that in the non-relativistic settings. Although the derivation is currently carried in the Minkowski spacetime, it should not be difficult to extend the derivation in a curved spacetime. The extended least action principle also provides interesting implications on the interpretation of quantum theory, which will be discussed in a separate report. 

The rest of the article is organized as follows. First, we briefly overview the least action principle for the classical scalar field, since it is the starting point of the quantum formulation. Second, we review the underlying assumptions for the extended least action principle and what should be adjusted to apply the principle in the case of scalar fields. In Section \ref{sec:QFT} we apply the principle recursively to analyze the dynamics of field fluctuations, then derive the uncertainty relation and the Schr\"{o}dinger equation for the wave functional.  The Schr\"{o}dinger equation is generalized in Section \ref{sec:GSE}. We then conclude the article after comprehensive discussions and comparisons to previous relevant research works.

\section{Classical Theory for Massive Scalar Fields}
\label{sec:classicalTheory}
This section briefly reviews the classical theory of scalar fields, the canonical transformation, and the Hamilton-Jacobi equation.
Consider a massive scalar field configuration $\phi$. Here we denote the coordinates for a four dimensional spacetime point $x$ either by $x=(x^{(0)}, x^{(i)})$ where $i=\{1, 2, 3\}$, or by $x=(t, \mathbf{x})$ where $\mathbf{x}$ is a spatial point. The field component at a spacetime point $x$ is denoted as $\phi_x=\phi(x)$. The Lagrangian density for the a massive scalar field is given by
\begin{equation}
    \label{LD}
    \begin{split}
        \mathcal{L} &= \frac{1}{2}[\partial_{\mu}\phi(x)]^2 - \frac{1}{2}m^2[\phi(x)]^2 \\
        &=\frac{1}{2}[\dot{\phi}(x)]^2 - \frac{1}{2}([\nabla\phi(x)]^2+m^2[\phi(x)]^2).
    \end{split}
\end{equation}
where $\mu=\{0, 1, 2, 3\}$ and the convention of Einstein summation is assumed. The first term $\frac{1}{2}[\dot{\phi}(x)]^2$ resembles the kinetic energy density in Newtonian mechanics, while the second term is the potential energy density and denoted as $V(\phi(x))$. The correspondent action functional is
\begin{equation}
    \label{action}
    A = \int d^4x\mathcal{L}.
\end{equation}
The momentum conjugate to the field is defined by
\begin{equation}
    \label{momentum}
    \pi(x) = \frac{\partial\mathcal{L}}{\partial(\partial_0\phi)}=\partial_0\phi(x) = \dot{\phi}(x).
\end{equation}
Applying the least action principle to minimize the action functional $S$, one obtains the Euler-Lagrange equation
\begin{equation}
    \label{KG}
    \partial_{\mu}\partial^{\mu}\phi + m^2\phi^2 = 0,
\end{equation}
which is the Klein-Gordon equation for the scalar field.

Variables $(\phi, \pi)$ form a pair of canonical variables, and the corresponding Hamiltonian is constructed by a Legendre transform of the Lagrangian~\cite{Long}
\begin{equation}
    \label{Hamiltonian}
    \begin{split}
    H[\phi,\pi] &= \int d^3\mathbf{x}\{\pi(x)\dot{\phi}(x) - \mathcal{L}\} \\
    &= \int d^3\mathbf{x} \{\frac{1}{2}[\dot{\phi}(x)]^2 + V\}.
    \end{split}
\end{equation}

Next we want to apply the canonical transformation technique in field theory. To do this, we will need to choose a foliation of the spacetime into a succession of spacetime hypersurfaces. Here we only consider the Minkowksi spacetime and it is natural to choose these to be the hypersurfaces $\Sigma_{t}$ of fixed $t$. The field configuration $\phi$ for $\Sigma_{t}$ can be understood as a vector with infinitely many components for each spatial point on the Cauchy hypersurface $\Sigma_t$ at time instance $t$ and denoted as $\phi_{t,\mathbf{x}}=\phi(t,\mathbf{x})$. For simplicity of notation, we will still denote $\phi(t,\mathbf{x})= \phi(x)$ for the rest of this paper, but the meaning of $\phi(x)$ should be understood as the field component $\phi_{\mathbf{x}}$ at each spatial point of the hypersurfaces $\Sigma_{t}$ at time instance $t$. In Appendix \ref{appendix:canonical}, we show that by an extended canonical transformation, the action functional of the field can be written as
\begin{equation}
    A_c = \int dt \{\frac{\partial S}{\partial t} + H[\phi, \pi]\},
\end{equation}
where $S[\phi, t]$ is a generation functional that satisfies the identity $\pi(x) = \delta S / \delta \phi(x)$. A special solution to the least action principle for the above action functional is $\partial S/\partial t + H = 0$. Substituting $H$ from (\ref{Hamiltonian}), we have
\begin{equation}
    \label{HJE}
    \frac{\partial S}{\partial t} + \int d^3\mathbf{x} \{\frac{1}{2}[\dot{\phi}(x)]^2 + V(\phi(x))\} = 0.
\end{equation}
Since $\dot{\phi}(x) = \pi(x) = \delta S / \delta \phi(x)$, the above equation can be rewritten as
\begin{equation}
    \label{HJE2}
    \frac{\partial S}{\partial t} + \int d^3\mathbf{x} \{\frac{1}{2}(\frac{\delta S}{\delta\phi(x)})^2 + V(\phi(x)\} = 0.
\end{equation}
This is the Hamilton-Jacobi equation for the scalar field that governs the evolution of the functional $S$ between the spacelike hypersurfaces. It is equivalent to the Klein-Gordon equation (\ref{KG}).

As also shown in Appendix \ref{appendix:canonical}, suppose the scalar field configuration $\phi$ follows a probability distribution, with probability density $\rho[\phi,t]$ for the hypersurface $\Sigma_t$, the average value of the action functional is,
\begin{equation}
    \label{extAction2}
    S_c = \int \mathcal{D}\phi dt \{\rho [\frac{\partial S}{\partial t} + \int d^3\mathbf{x} \{\frac{1}{2}(\frac{\delta S}{\delta\phi(x)})^2 + V(\phi(x)]\}.
\end{equation}
Note that $S_c$ and $S$ are different functional, where $S_c$ can be considered as the ensemble average of classical action functional and $S$ is a generation functional introduced in an extended canonical transformation that satisfied $\pi(x) = \delta S / \delta \phi(x)$. Now we consider the generalized canonical pair as $(\rho, S)$, and apply the least action principle on the action functional defined in (\ref{extAction2}). Variation of $S_c$ over $\rho$ leads to (\ref{HJE2}), and variation of $S_c$ over $S$ gives
\begin{equation}
    \label{contEq}
    \frac{\partial \rho}{\partial t} + \int \frac{\delta}{\delta\phi(x)}(\rho\frac{\delta S}{\delta\phi(x)}) d^3\mathbf{x} = 0,
\end{equation}
which is the continuity equation for the probability density. Both equations (\ref{HJE2}) and (\ref{contEq}) determine the dynamics of the classical scalar field ensemble, and they are obtained by applying the least action principle based on the action functional $S_c$ defined in (\ref{extAction2}).

\section{\replaced{The Extended Least Action Principle}{Principle of Least Observability}}
\label{sec:LIP}
Ref.~\cite{Yang2023} shows that the least action principle in classical mechanics can be extended to derive quantum formulation by factoring in the following two assumptions. 
\begin{displayquote}
\emph{Assumption 1 -- A quantum system experiences vacuum fluctuations constantly. The fluctuations are local and completely random.}
\end{displayquote}
\begin{displayquote}
\emph{Assumption 2 -- There is a lower limit to the amount of action that a physical system needs to exhibit in order to be observable. This basic discrete unit of action effort is given by $\hbar/2$ where $\hbar$ is the Planck constant.}
\end{displayquote}

The first assumption is generally accepted in mainstream quantum mechanics, which is responsible for the intrinsic randomness of the dynamics of a quantum object. Locality of vacuum fluctuation is assumed, and it implies that for a composite system, the fluctuation of each subsystem is independent of each other. 

The justifications of the second assumption is explained in detail in Section II of Ref.~\cite{Yang2023}. Historically the Planck constant was first introduced to show that the energy of radiation from a black body is discrete. One can consider the discrete energy unit as the smallest unit to be distinguished, or detected, in the black body radiation phenomenon. In general, it is understood that Planck constant is associated with the discreteness of certain observable in quantum mechanics. Here, we just interpret the Planck constant from an information measure point of view. Essentially, what we assume is that there is a lower limit to the amount of action that the physical system needs to exhibit in order to be observable or distinguishable in potential observation, and such a unit of action is defined by the Planck constant. 

\added{Making use of this understanding of the Planck constant inversely provides us a new way to calculate the additional action due to vacuum fluctuations. That is, even though we do not know the physical details of vacuum fluctuations, the vacuum fluctuations manifest themselves via a discrete action unit determined by the Planck constant as an observable information unit. If we are able to define an information metric that quantifies the amount of observable information manifested by vacuum fluctuations, we can then multiply the metric with the Planck constant to obtain the action associated with vacuum fluctuations. Then, the challenge to calculate the additional action due to vacuum fluctuation is converted to define a proper new information metric $I_f$, which measures the additional distinguishable, hence observable, information exhibited due to vacuum fluctuations. Even though we do not know the physical details of vacuum fluctuations (except that as Assumption 1 states, these vacuum fluctuations are completely random and local), the problem becomes less challenged since there are information-theoretic tools available. The first step is to assign a transition probability distribution due to vacuum fluctuation for an infinitesimal time step at each position along the classical trajectory. The distinguishability of vacuum fluctuation then can be defined as the information distance between the transition probability distribution and a uniform probability distribution. Uniform probability distribution is chosen here as reference to reflect the complete randomness of vacuum fluctuations. In information theory, the common information metric to measure the information distance between two probability distributions is relative entropy. Relative entropy is more fundamental to Shannon entropy since the latter is just a special case of relative entropy when the reference probability distribution is a uniform distribution. But there is a more important reason to use relative entropy. As shown in later sections, when we consider the dynamics of the system for an accumulated time period, we assume the initial position is unknown but is given by a probability distribution. This probability distribution can be defined along the position of classical trajectory without vacuum fluctuations, or with vacuum fluctuations. The information distance between the two probability distributions gives the additional distinguishability due to vacuum fluctuations. It is again measured by a relative entropy. Thus, relative entropy is a powerful tool allowing us to extract meaningful information about the dynamic effects of vacuum fluctuations. Concrete form of $I_f$ will be defined later as a functional of Kullback-Leibler divergence $D_{KL}$, $I_f:=f(D_{KL})$, where $D_{KL}$ measures the information distances of different probability distributions caused by vacuum fluctuations. Thus, the total action from classical path and vacuum fluctuation is}
\begin{equation}
\label{totalAction}
    S_t = S_c + \frac{\hbar}{2}I_f,
\end{equation}
\added{where $S_c$ is the classical action. Non-relativistic quantum theory can be derived~\cite{Yang2023} through a variation approach to minimize such a functional quantity, $\delta S_t=0$. When $\hbar \to 0$, $S_t=S_c$. Minimizing $S_t$ is then equivalent to minimizing $S_c$, resulting in Newton's laws in classical mechanics. However, in quantum mechanics, $\hbar \ne 0$, the contribution from $I_f$ must be included when minimizing the total action. We can see $I_f$ is where the quantum behavior of a system comes from. These ideas can be condensed as
\begin{displayquote}
\emph{\textbf{Extended Principle of Least Action} -- The law of physical dynamics for a quantum system tends to exhibit as little as possible the action functional defined in (\ref{totalAction}).}
\end{displayquote}}

\deleted{The existing of the Planck constant as a unit of action for the physical system to exhibit in order to be observable allows us to ask the question: Given certain amount of action $S_c$ from its motion along a classical trajectory, how much observability does the system exhibit from its dynamics? According to assumption 2, this is calculated as $I_p = 2S_c/\hbar$. $I_p$ is not a conventional information metric but it has clear meaning about a piece of physical information. That is, it can be considered as the amount of observable information measured in the unit of $\hbar/2$. This step of converting $S_c$ into $I_p$ appears trivial mathematically, but conceptually it is not. It recasts the least action principle into a least observability principle, and shifts the working language to be information related. Thus, $I_p$ can be paired with additional information metrics due to vacuum fluctuations. To measure the degree of observability due to vacuum fluctuations, a new information metric $I_f$ is introduced. $I_f$ is defined as a metric to measure the additional distinguishable, hence observable, information exhibited due to vacuum fluctuations. $I_f$ will be defined as a functional of Kullback-Leibler divergence $D_{KL}$, $I_f:=f(D_{KL})$, where $D_{KL}$ measures the information distances of different probability distributions caused by vacuum fluctuations. Thus, the total degree of observability due to both classical trajectory and vacuum fluctuation is}
\deleted{
Non-relativistic quantum theory can be derived through a variation method that demands $I$ is stationary, that is, $\delta I=0$. When $\hbar \to 0$, the observability due to classical path $I_p \to \infty$. Thus, the system can be observed with infinite accuracy, and any finite amount of $I_f$ can be ignored. Minimizing $I$ is then equivalent to minimizing $S_c$, resulting in the dynamics laws of classical mechanics. However, in quantum mechanics, according to the Assumption 2, the action to exhibit observability is discrete so that $\hbar \ne 0$, $I_p$ is finite. This means there is only finite amount of observable information available. The contribution from $I_f$ can be comparable to $I_f$ and therefore must be included when minimizing the total amount of observable information. These ideas can be condensed as}

Now we want to apply this principle to the scalar field and derive the quantum scalar field theory. Assumption 1 needs to be slightly modified, since in the field theory, one does not deal with a physical object. Instead, we are dealing with the field configuration. Assumption 1 is restated as 
\begin{displayquote}
\emph{Assumption 1a -- There are constant fluctuations in the field configurations. The fluctuations are completely random, and local.}
\end{displayquote}
It is not our intention here to investigate the origin, or establish a physical model, of such field fluctuations. Instead, we make a minimal number of assumptions on the underlying physical model, only enough so that we can apply the variation principle based on minimizing the total action.

Assumption 2 is unchanged for quantum field theory. The action of the classical scalar field $S_c$ is given by (\ref{action}), or (\ref{extAction2}). Similarly, the metrics to measure the additional distinguishable information exhibited due to field fluctuations, is defined as a functional of Kullback-Leibler divergence $D_{KL}$, $I_f:=f(D_{KL})$, where $D_{KL}$ measures the information distances of different probability distributions caused by field fluctuations. Thus, the total action due to both classical field dynamics and field fluctuation is given by the same equation as (\ref{totalAction}). Quantum field theory can be derived through a variation method to minimize such a functional quantity, $\delta S_t=0$. 

\added{Alternatively, we can interpret the extended least action principle more from an information perspective by rewriting (\ref{totalAction}) as 
\begin{equation}
\label{totalInfo}
    I_t =\frac{2}{\hbar} S_c + I_f,
\end{equation}
where $I_t=2S_t/\hbar$. Denote $I_p=2S_c/\hbar$, which measures the amount of $S_c$ using the discrete unit $\hbar/2$. $I_p$ is not a conventional information metric but can be considered carrying meaningful physical information about the observability of the classical field. More discussion on the meaning of observability is provided later in Section \ref{sec:discussion}. Similarly, $I_f$ measures the distinguishable information of the probability distributions with and without field fluctuations. Thus, $I_t$ is the total observable information. With (\ref{totalInfo}), the extended least action principle can be restated as\footnote{The term observability is not related to the concept of observable in traditional quantum physics since it is not associated with a Hermitian operator. Also, one should not confuse the term with the same terminology in system control theory.} 
\begin{displayquote}
\emph{\textbf{Principle of Least Observability} -- The law of physical dynamics for a quantum field tends to exhibit as little as possible the observable information defined in (\ref{totalInfo}).}
\end{displayquote}
Mathematically, there is no difference between (\ref{totalAction}) and (\ref{totalInfo}) when applying the variation principle to derive the laws of field dynamics. The form of (\ref{totalAction}) in terms of actions appears more familiar in the physics community. However, The form of (\ref{totalInfo}) in terms of observability seems conceptually more general. We will leave the exact interpretations of the principle alone and use the two interpretations interchangeable in this paper. The key point to remember is that the Planck constant connects the physical action to metrics related to observable information in either interpretation.}
\deleted{When $\hbar \to 0$, the observability due to classical field $2S_c/\hbar \to \infty$, any finite amount of $I_f$ can be ignored. Minimizing $I$ is then equivalent to minimizing $S_c$, resulting in the dynamics laws of classical fields. However, in quantum field theory, the action to exhibit observability is discrete so that $\hbar \ne 0$, $2S_c/\hbar$ is finite. Therefore, the contribution from $I_f$ must be included when minimizing the total degree of observability. }

Next we will show that by applying the variational principle to minimize the action functional defined in (\ref{totalAction}), we can obtain the uncertainty relation and the Schr\"{o}dinger equation of the wave functional for the scalar field, which are the basic formulation of the quantum scalar field.


\section{Quantum Theory for Massive Scalar Fields}
\label{sec:QFT}
\subsection{Field Fluctuations and Uncertainity Relation}
\label{sec:shorttime}
First we consider the field fluctuations in an equal times hyper-surfaces for an infinitesimal time internal $\Delta t$. At a given time $t\to t+\Delta t$ in the hyper-surface $\Sigma_t$, the field configuration fluctuates randomly, $\phi\to \phi + \omega$, where $\omega=\Delta\phi$ is the change of field configuration due to random fluctuations. Define the probability for the field configuration to transition from $\phi$ to $\phi + \omega$ as $p[\phi + \omega|\phi]\mathcal{D}\omega$. The expectation value of classical action over all possible field fluctuations is $S_c=\int p[\phi + \omega|\phi]\mathcal{L}d^3\mathbf{x}\mathcal{D}\omega dt$ where $\mathcal{L}$ is given by (\ref{LD}) for a scalar field. For an infinitesimal time internal $\Delta t$, one can approximate $\dot\phi=\Delta\phi/\Delta t=\omega/\Delta t$. The classical action for the infinitesimal time internal $\Delta t$ is approximately given by 
\begin{equation}
\label{action1}
    S_c=\int p[\phi + \omega|\phi]\mathcal{D}\omega\int_{\Sigma_t}\{\frac{[\omega(x)]^2}{2\Delta t}+V(\phi(x))\Delta t\}d^3\mathbf{x}.
\end{equation}
The information metrics $I_f$ is supposed to capture the additional revelation of information due to field fluctuations in $\Sigma_t$. Thus, it is naturally defined as a relative entropy, or more specifically, the Kullback–Leibler divergence, to measure the information distance between $p[\phi + \omega|\phi]$ and some prior probability distribution. Since the field fluctuations are completely random, it is intuitive to assume the prior distribution with maximal ignorance~\cite{Caticha2019, Jaynes}. That is, the prior probability distribution is a uniform distribution $\sigma$. 
\begin{align*}
    I_f  &:= D_{KL}(p[\phi + \omega|\phi]|| \sigma) \\
    &= \int p[\phi + \omega|\phi]ln[p[\phi + \omega|\phi]/\sigma]\mathcal{D}\omega.
\end{align*}
Combined with (\ref{action1}), the total action functional defined in (\ref{totalAction}) is
\begin{align*}
    S_t = &\int p[\phi + \omega|\phi]\mathcal{D}\omega\int (\frac{[\omega(x)]^2}{2\Delta t} + V(\phi(x)) \Delta t)d^3\mathbf{x} \\
        &+ \frac{\hbar}{2}\int p[\phi + \omega|\phi]ln[p[\phi + \omega|\phi]/\sigma]\mathcal{D}\omega.
\end{align*}
Taking the variation $\delta S_t = 0$ with respect to $p$ gives 
\begin{equation}
    \delta S_t = \frac{\hbar}{2}\int \{ \int (\frac{[\omega(x)]^2}{\hbar\Delta t} + \frac{2V \Delta t}{\hbar})d^3\mathbf{x}+ln\frac{p}{\sigma} +1)\}\delta p \mathcal{D}\omega = 0.
\end{equation}
Since $\delta p$ is arbitrary, one must have 
\begin{equation*}
    \int ([\omega(x)]^2 + 2V (\Delta t)^2]d^3\mathbf{x}+\hbar\Delta t(ln\frac{p}{\sigma} +1)=0.
\end{equation*}
When $\Delta t$ is infinitesimally small, we can ignore the higher order term with $(\Delta t)^2$, and obtain the solution for $p$ as
\begin{equation}
\label{transP}
\begin{split}
    p[\phi + \omega|\phi] &= \sigma e^{-\frac{1}{\hbar\Delta t}\int[\omega(x)]^2 d^3\mathbf{x} - 1} \\
    &= \frac{1}{Z}e^{-\frac{1}{\hbar\Delta t}\int[\omega(x)]^2 d^3\mathbf{x}},
    \end{split}
\end{equation}
where $Z$ is a normalization factor that absorbs factor $\sigma e^{-1}$. Equation (\ref{transP}) shows that the transition probability density is a Gaussian-like distribution. It is independent of $\phi$ and can be simply denoted as $p[\omega]$. Clearly, the expectation value of $\omega(x)$ is
\begin{equation}
    \label{avergareW}
    \langle \omega(x)\rangle = \int p[\omega] \omega(x) \mathcal{D}\omega = 0.
\end{equation}
We also want to evaluate the expectation value field fluctuations at two spatial points in hypersurface $\Sigma_t$, $x=(t, \mathbf{x})$ and $x'=(t, \mathbf{x'})$,
\begin{equation}
    \label{corrW}
    \langle \omega(x)\omega(x')\rangle = \int p[\omega] \omega(x) \omega(x')\mathcal{D}\omega.
\end{equation}
In Appendix \ref{appendix:varW}, we verify that 
\begin{equation}
    \label{varW}
    \langle \omega(x)\omega(x')\rangle = \frac{\hbar\Delta t}{2}\delta (\mathbf{x}-\mathbf{x'}),
\end{equation}

Recall that $\omega = \Delta\phi$, and $\pi=\dot\phi=\Delta\phi/\Delta t=\omega/\Delta t$. Since $\langle\omega\rangle=0$, one has $\langle \pi\rangle=\langle\omega\rangle/\Delta t = 0$ as well. Thus, $\Delta\pi = \pi - \langle \pi\rangle = \pi = \omega/\Delta t$, we re-arrange (\ref{varW}) as
\begin{equation}
    \label{varW3}
    \langle \Delta\phi(x)\Delta\pi(x')\rangle  = \frac{\hbar}{2}\delta (\mathbf{x}-\mathbf{x'}).
\end{equation}
Applying the Cauchy–Schwarz inequality we get
\begin{equation}
    \label{uncertainty}
    \langle \Delta\phi(x)\rangle\langle\Delta\pi(x')\rangle \ge \langle\Delta\phi(x)\Delta\pi(x')\rangle =\frac{\hbar}{2}\delta (\mathbf{x}-\mathbf{x'}).
\end{equation}
But comparing with the $\delta$-function in the right hand side of (\ref{uncertainty}) appears inappropriate. Instead, we introduce a pair of positive spatial test functions $f(\mathbf{x}), g(\mathbf{x}): \mathbb{R}^3\to\mathbb{R}^+$, and define
\begin{equation}
    \langle \omega(f)\omega(g)\rangle = \int p[\omega] \{\int_{\Sigma_t}\omega(x) f(\mathbf{x}) \omega(x')g(\mathbf{x'})d\mathbf{x}d\mathbf{x'}\}\mathcal{D}\omega.
\end{equation}
Repeating the similar calculations from (\ref{varW}) to (\ref{uncertainty}), we can obtain
\begin{equation}
    \label{uncertainty2}
    \langle \Delta\phi(f)\rangle\langle\Delta\pi(g)\rangle \ge \frac{\hbar}{2}\langle f | g \rangle,
\end{equation}
where $\langle f | g \rangle = \int_{\Sigma_t}f(\mathbf{x})g(\mathbf{x})d\mathbf{x}$. This is the uncertainty relation between the field variable $\phi$ and its conjugate momentum variable $\pi$ for the scalar fields.

\subsection{Derivation of The Schr\"{o}dinger Equation for the Wave Functional} 
\label{sec:SE}
We now turn to the field dynamics for a period of time from $t_A\to t_B$. As described earlier, the spacetime during the time duration $t_A\to t_B$ is sliced into a succession of $N$ Cauchy hypersurfaces $\Sigma_{t_i}$, where $t_i \in \{t_0=t_A, \ldots, t_i, \ldots, t_{N-1}=t_B\}$, and each time step is an infinitesimal period $\Delta t$. The field configuration for each $\Sigma_{t_i}$ is denoted as $\phi(t_i)$, which has infinite number of components, labeled as $\phi_{\mathbf{x}}(t_i)=\phi(\mathbf{x}, t_i)$, for each spatial point in $\Sigma_{t_i}$. Without considering the random field fluctuation, the dynamics of the field configuration is governed by the Hamilton-Jacobi equation (\ref{HJE2}). Furthermore, we consider an ensemble of field configurations for hypersurface $\Sigma_{t_i}$ that follow a probability density\footnote{The notation $\rho[\phi, t_i]$ is legitimate since in this case $\phi$ describes the field configuration for the equal time hypersurface $\Sigma_{t_i}$.} $\rho_{t_i}[\phi] = \rho[\phi, t_i]$ which follows the continuity equation (\ref{contEq}). As shown in Section \ref{sec:classicalTheory}, both the Hamilton-Jacobi equation and the continuity equation can be derived through variation over the classical action functional $S_c$, as defined in (\ref{extAction2}), with respect to $\rho$ and $S$, respectively. 

To apply the extended least action principle, first we compute the action from the dynamics of the classical field ensemble as defined in (\ref{extAction2}). Next we need to define the information metrics for the field fluctuations, $I_f$. For each new field configuration $\phi+\omega$ due to the field fluctuations, there is a new probability density $\rho[\phi+\omega, t_i]$. We need a proper metrics to measure the additional revelation of observable information due to the field fluctuations on top of the classical field dynamics. The proper measure of this distinction is the information distance between $\rho[\phi, t_i]$ and $\rho[\phi+\omega, t_i]$. A natural choice of such information measure is the relative entropy $D_{KL}(\rho[\phi, t_i] || \rho[\phi+\omega, t_i])$. Moreover, we need to consider the contributions for all possible $\omega$. Thus, we take the expectation value of $D_{KL}$ over $\omega$, denoted as $\langle\cdot\rangle_{\omega}$. Then, the contribution of distinguishable information due to field fluctuations for hypersurce $\Sigma_{t_i}$ is $\langle D_{KL}(\rho[\phi, t_i] || \rho[\phi+\omega, t_i])\rangle_{\omega}$. Finally, we sum up the contributions from all hypersurfaces, lead to the definition of information metrics
\begin{align}
\label{DLDivergence}
    I_f &:= \sum_{i=0}^{N-1}\langle D_{KL}(\rho[\phi, t_i] || \rho[\phi+\omega, t_i])\rangle_{\omega} \\
    &=\sum_{i=0}^{N-1}\int \mathcal{D}\omega p[\omega] \int \mathcal{D}\phi \rho [\phi, t_i]ln \frac{\rho[\phi, t_i]}{\rho [\phi+\omega, t_i]}.
\end{align}
Notice that $p[\omega]$ is a Gaussian-like distribution given in (\ref{transP}). When $\Delta t$ is small, only small fluctuations $\omega$ will contribute to $I_f$. As shown in Appendix \ref{appendix:SE}, when $\Delta t\to 0$, $I_f$ turns out to be
\begin{equation}
\label{FisherInfo}
    I_f = \frac{\hbar}{4}\int \frac{1}{\rho[\phi, t]}(\frac{\delta\rho[\phi, t]}{\delta\phi(x)})^2 d^3\mathbf{x} \mathcal{D}\phi dt .
\end{equation}
Eq. (\ref{FisherInfo}) is analogous to the Fisher information for the probability density~\cite{Yang2023, FriedenBook} in non-relativistic quantum mechanics. Some literature directly adds such Fisher information term in the variation method as a postulate to derive the Schr\"{o}dinger equation~\cite{Caticha2014, Ipek2021}. But (\ref{FisherInfo}) bears much more physical significance than Fisher information. First, it shows that $I_f$ is proportional to $\hbar$. This is not trivial because it avoids introducing additional arbitrary constants for the subsequent derivation of the Schr\"{o}dinger equation. More importantly, defining $I_f$ using relative entropy opens up new results that cannot be obtained if $I_f$ is defined using Fisher information, because there are other generic forms of relative entropy such as R\'{e}nyi divergence or Tsallis divergence. As will be seen later, by replacing the Kullback–Leibler divergence with R\'{e}nyi divergence, one will obtain a family of generalized Schr\"{o}dinger equations. 

Together with (\ref{extAction2}), (\ref{FisherInfo}), and (\ref{totalAction}), the total action functional is
\begin{equation}
    \label{totalDist}
    \begin{split} 
    S_t =& \int\rho\{\frac{\partial S}{\partial t} + \int [\frac{1}{2}(\frac{\delta S}{\delta\phi(x)})^2 + V(\phi(x)) \\
    & + \frac{\hbar^2}{8}(\frac{1}{\rho}\frac{\delta\rho}{\delta\phi(x)})^2 ]d^3\mathbf{x}\}\mathcal{D}\phi dt.
    \end{split}
\end{equation}
Variation of $S_t$ with respect to $S$ gives the same continuity equation (\ref{contEq}), while variation with respect to $\rho$ leads to (see Appendix \ref{appendix:SE})
\begin{equation}
\label{QHJ}
\begin{split}
    \frac{\partial S}{\partial t} =- \int\{\frac{1}{2}(\frac{\delta S}{\delta\phi(x)})^2 + V(\phi(x)) - \frac{\hbar^2}{2R}\frac{\delta^2 R}{\delta\phi^2(x)} \}d^3\mathbf{x},
    \end{split}
\end{equation}
where $R[\phi, t]=\sqrt{\rho[\phi, t]}$. The last term in the R.H.S. of (\ref{QHJ}) is the scalar field equivalence of the Bohm quantum potential~\cite{Bohm1952}. In non-relativistic quantum mechanics, the Bohm potential is considered responsible for the non-locality phenomenon in quantum mechanics~\cite{Bohm2}. Its origin is mysterious. Here we show that it originates from the information metrics related to relative entropy, $I_f$. 

Defined a complex functional $\Psi[\phi,t]=R[\phi, t]e^{iS[\phi, t]/\hbar}$, the continuity equation and the extended Hamilton-Jacobi equation (\ref{QHJ}) can be combined into a single functional derivative equation (see Appendix \ref{appendix:SE}),
\begin{equation}
    \label{SE}
    i\hbar\frac{\partial\Psi[\phi, t]}{\partial t} = \{\int[-\frac{\hbar^2}{2}\frac{\delta^2}{\delta\phi^2(x)} + V(\phi(x))]d^3\mathbf{x}\}\Psi[\phi, t].
\end{equation}
This is the Schr\"{o}dinger equation for the wave functional $\Psi[\phi,t]$ with Hamiltonian operator 
\begin{equation}
    \label{operatorH}
    \hat{\mathcal{H}} = -\frac{\hbar^2}{2}\frac{\delta^2}{\delta\phi^2(x)} + V(\phi(x)).
\end{equation}
It governs the evolution of wave functional $\Psi[\phi,t]$ between hypersurfaces $\Sigma_t$. The potential density in (\ref{SE}), for the massive scalar field, is given in (\ref{LD}) as $V(\phi(x))=\frac{1}{2}([\nabla\phi(x)]^2+m^2[\phi(x)]^2)$. But it can be generalized to be
\begin{equation}
    \label{potentialDensity}
    \begin{split} 
    V(\phi(x))=&\frac{1}{2}[\nabla\phi(x)]^2+\frac{m^2}{2}[\phi(x)]^2 \\
    &+ \lambda[\phi(x)]^3 + \lambda'[\phi(x)]^4+ \ldots
    \end{split}
\end{equation}
where the coefficients $\lambda$, $\lambda'$, represent mass and other coupling constants. Once the Schr\"{o}dinger equation for the wave functional $\Psi[\phi,t]$ is obtained, other standard results follow, such as the solutions for the wave functional and the energy of the ground state and excited state~\cite{Long}.

In summary, by recursively applying the same extended least action principle in two steps, we recover the uncertainty relation and the Schr\"{o}dinger representations of the standard relativistic quantum theory of scalar field~\cite{Long, Jackiw}. In the first step, we analyze the dynamics of field fluctuations in a hypersurface $\Sigma_t$ for a short period of time interval $\Delta t$, and obtain the transitional probability density due to field fluctuations; In the second step, we apply the principle for a cumulative time period to obtain the dynamics laws that govern the evolutions of $\rho$ and $S$ between the hypersurfaces. The applicability of the same principle in both steps shows the consistency and simplicity of the theory, although the forms of Lagrangian density are different in each step. In the first step, the Lagrangian density $\mathcal{L}$ is given by (\ref{LD}), while in the second step, we use a different form of Lagrangian density $\mathcal{L}^\prime = \rho(\partial S/\partial t + H)$. As shown in Appendix \ref{appendix:canonical}, $\mathcal{L}$ and $\mathcal{L}^\prime$ are related through an extended canonical transformation. The choice of Lagrangian $\mathcal{L}$ or $\mathcal{L}^\prime$ does not affect the out of variation procedure, that is, the form of Legendre's equations. We choose $\mathcal{L}^\prime$ as the Lagrangian density in the second step in order to use the pair of functional $(\rho, S)$ in the subsequent variation procedure.  

\added{It is important to point out that the derivation of (\ref{SE}) depends on a particular foliation of the Minkowski spacetime. Therefore, the theoretical framework presented here treats time parameter differently and it is not obvious if the theory is Lorentz invariance. The issue is extensively studied in \cite{Ipek2019,Ipek2021}, and the answer is that the theory is still fully relativistic. This is because using the resulting Hamiltonian operator $\hat{\mathcal{H}}$ given by (\ref{operatorH}) and (\ref{potentialDensity}), one can identify the generators for translation and rotation operations for both time-like and spatial-like directions, and these generators satisfy the Poincar\'{e} algebra\cite{Ipek2021}. Although the theory singles out a particular time parameter for use through the foliation of spacetime, the Poincar\'{e} algebra guarantees that the resulting dynamical evolution is fully relativistic. This is because satisfying this algebra guarantees that one can construct a Poincar\'{e} covariant stress-energy tensor for the dynamical variables\footnote{Note that even though the way we derive the Schr\"{o}dinger equation for the wave functional is different from that in \cite{Ipek2019,Ipek2021}, once both theories agree on the Schr\"{o}dinger equation and the Hamiltonian operator, the procedure to identify the generators that satisfy the Poincare algebra is the same. Thus, the discussions in \cite{Ipek2019,Ipek2021} regarding the compliance to relativistic theory is applicable here.}. 
}

\section{The Generalized Schr\"{o}dinger Equation for the Wave Functional}
\label{sec:GSE}
As mentioned earlier, by relaxing the definition of the information metrics $I_f$, one can generalize the Schr\"{o}dinger equation for the wave functional. The term $I_f$ is supposed to capture the additional distinguishability exhibited by the field fluctuations, and is defined in (\ref{DLDivergence}) as the summation of the expectation values of Kullback–Leibler divergence between $\rho[\phi,t]$ and $\rho[\phi+\omega,t]$. However, there are more generic definitions of relative entropy, such as the R\'{e}nyi divergence~\cite{Renyi, Erven2014}. From an information theoretic point of view, it is legitimate to consider alternative definitions of relative entropy. Suppose we define $I_f$ based on R\'{e}nyi divergence,
\begin{align}
\label{RDivergence}
    I_f^{\alpha} &:= \sum_{i=0}^{N-1}\langle D^{\alpha}_R(\rho[\phi, t_i] || \rho[\phi+\omega, t_i])\rangle_{\omega} \\
    &=\sum_{i=0}^{N-1}\int  \mathcal{D}\omega p[\omega] \frac{1}{\alpha-1}ln (\int   \mathcal{D}\phi\frac{\rho^{\alpha}[\phi, t_i]}{\rho^{\alpha-1}[\phi+\omega, t_i]}).
\end{align}
Parameter $\alpha \in (0,1)\cup(1, \infty)$ is called the order of R\'{e}nyi divergence. When $\alpha\to 1$, $I_f^{\alpha}$ converges to $I_f$ as defined in (\ref{DLDivergence}). In Appendix \ref{appendix:RE}, we show that using $I_f^{\alpha}$ and following the same variation principle, we arrive at a similar extended Hamilton-Jacobi equation as (\ref{QHJ}),
\begin{equation}
\label{RHJ}
\begin{split}
    \frac{\partial S}{\partial t} =- \int\{\frac{1}{2}(\frac{\delta S}{\delta\phi(x)})^2 + V(\phi(x)) - \frac{\alpha\hbar^2}{2R}\frac{\delta^2 R}{\delta\phi^2(x)} \}d^3\mathbf{x},
    \end{split}
\end{equation}
with an additional coefficient $\alpha$ appearing in the Bohm quantum potential term. Defined a complex functional $\Psi_{\alpha}[\phi,t]=R[\phi, t]e^{iS[\phi, t]/\sqrt{\alpha}\hbar}$, the continuity equation and the extended Hamilton-Jacobi equation (\ref{RHJ}) can be combined into an equation similar to the Schr\"{o}dinger equation (see Appendix \ref{appendix:RE}),
\begin{equation}
    \label{SE2}
    i\sqrt{\alpha}\hbar\frac{\partial\Psi_{\alpha}[\phi,t]}{\partial t} = \{\int[-\frac{\alpha\hbar^2}{2}\frac{\delta^2}{\delta\phi^2(x)} + V(\phi(x))]d^3\mathbf{x}\}\Psi_{\alpha}[\phi, t].
\end{equation}
When $\alpha=1$, the regular Schr\"{o}dinger equation of wave functional (\ref{SE}) is recovered, as expected. Equation (\ref{SE2}) gives a family of linear equations for each order of R\'{e}nyi divergence. 

As observed in Appendix \ref{appendix:RE}, if we define $\hbar_{\alpha}= \sqrt{\alpha}\hbar$, then $\Psi_{\alpha}[\phi,t]=R[\phi, t]e^{iS[\phi, t]/\hbar_{\alpha}}$, and (\ref{SE2}) becomes the same form of the regular Schr\"{o}dinger equation (\ref{SE}) but with replacement of $\hbar$ to $\hbar_{\alpha}$. It is as if there is an intrinsic relation between the order of R\'{e}nyi divergence $\alpha$ and the Plank constant $\hbar$. This remains to be investigated further. 
On the other hand, if the wavefunction is defined as usual without the factor $\sqrt{\alpha}$, $\Psi[\phi,t]=R[\phi, t]e^{iS[\phi, t]/\hbar}$, it will result in a nonlinear Schr\"{o}dinger equation for the wave functional. This implies that the linearity of Schr\"{o}dinger equation depends on how the wave functional is defined from the pair of real functional $(\rho, S)$. 

We also want to point out that $I_f^{\alpha}$ can be defined using Tsallis divergence~\cite{Tsallis, Nielsen2011} as well, instead of using the R\'{e}nyi divergence,
\begin{align}
\label{TDivergence}
\begin{split}
    I_f^{\alpha} &:= \sum_{i=0}^{N-1}\langle D^{\alpha}_T(\rho[\phi, t_i] || \rho[\phi+\omega, t_i])\rangle_{\omega} \\
    &=\sum_{i=0}^{N-1}\int  \mathcal{D}\omega p[\omega] \frac{1}{\alpha-1}\{\int  \mathcal{D}\phi\frac{\rho^{\alpha}[\phi, t_i]}{\rho^{\alpha-1}[\phi+\omega, t_i]} -1\}.
\end{split}
\end{align}
When $\Delta t\to 0$, it can be shown that the $I_f^\alpha$ defined above converges into the same form as (\ref{I_f4}). Hence it results in the same generalized Schr\"{o}dinger equation (\ref{SE2}).

\section{Discussion and conclusions} 
\label{sec:discussion}

\subsection{Alternative Formulation of the Extended Least Action Principle}
\deleted{Rewriting Eq.(\ref{totalInfo}) as $J = S_c + (\hbar/2)I_f$, and performing the same variation procedure will give the same Schr\"{o}dinger equation of the wave functional for the scalar field. However, the physical interpretation is different. One would need to consider $(\hbar/2)I_f$ as additional action due to field fluctuations. In other words, one would compute the amount of distinguishability information $I_f$ first, then apply Assumption 2 to convert the information quantity into action quantity. It is mathematically equivalent to Eq. (\ref{totalInfo}). But the question is how such action effort is physically realized. To answer this question, it requires a physical model for the field fluctuations at the sub-quantum level. This is challenged and beyond the scope of this paper.}
\added{We mention in Section \ref{sec:LIP}) that the extended least action principle can be restated as the principle of least observability by interpreting $I_p=2S_c/\hbar$ as the observable information of the classical trajectory. $I_p$ is not a conventional information metric but can be considered carrying meaningful physical information. To see this connection, recall that the classical action is defined as an integral of the Lagrangian over a period of time along a path trajectory of a classical object. There are two aspects to understanding the action functional. In classical mechanics, the path trajectory can be traced, measured, or observed. Given two fixed end points, the longer the path trajectory, the larger the value of the action. It indicates 1.) the more dynamic effort the system exhibits; and 2.) the easier to trace the path and distinguish the object from the background reference frame, or in other words, the more physical information available for potential observation. Thus, action $S_c$ not only quantifies the dynamic effort of the system, but also is associated with the detectability, or observability, of the physical object during the dynamics along the path. In classical mechanics, we focus on the first aspect via the least action principle, and derive the law of dynamics from minimizing the action effort. The second aspect is not useful since we cannot quantify the intuition that $S$ is associated with the observability of the physical object. One reason is that there is no natural unit of action to convert $S$ into an information related metric. The introduction of the Planck constant in Assumption 2 helps to quantify this intuition.}

Alternatively, we can interpret the least observability principle based on Eq. (\ref{totalInfo}) as minimizing $I_f$ with the constraint of $S_c$ being a constant, and $\hbar/2$ simply being a Lagrangian multiplier for such a constraint. Again, mathematically, it is an equivalent formulation. In that case, Assumption 2 is not needed. Instead it will be replaced by the assumption that the classical action functional $S_c$ is a constant with respect to variations on $\rho$ and $S$. But such an assumption needs sound justification. Which assumption to use depends on which choice is more physically intuitive. We believe that the least observability principle based on Assumption 2, where the Planck constant defines the discrete unit of action effort to exhibit observable information, gives more intuitive physical meaning of the mathematical formulation and without the need of a physical model for the field fluctuations.

\subsection{Comparisons with Relevant Research Works}
The Schr\"{o}dinger equation for the wave functional of scalar fields is typically introduced as a postulate~\cite{Long, Jackiw} instead of derived from a first principle. An impressive attempts to derive it from the entropic dynamics approach can be found in Ref.~\cite{Caticha2014, Ipek2021}. The entropic dynamic approach bears some similarity with the theory presented in this work. For instance, the formulations are carried out with two steps, an infinitesimal time step and a cumulative time period. It also aims to derive the physical dynamics by extremizing information quantity such as the relative entropy. However, the entropic dynamics approach relies on another postulate on energy conservation to complete the derivation of the Schr\"{o}dinger equation. The theory presented in this paper, on the other hand, has the advantage of simplicity since it recursively applies the same least observability principle in both infinitesimal time step and cumulative time period. The entropic dynamics approach also requires several seemingly arbitrary constants in their formulations, while we only need the Planck constant $\hbar$ and its meaning is clearly given in Assumption 2. We clearly show the the Bohm quantum potential term in (\ref{QHJ}) is originated from the information metrics of field fluctuations $I_f$, while ~\cite{Caticha2014, Ipek2019, Ipek2021} justify it from information geometry perspective. The advantages of our approach have two fold. First, it is far more conceptually clear to define $I_f$ as expectation value of relative entropy between different probability distribution due to field fluctuations. There is clear physical meaning associated with $I_f$. Second, we show that by using the general definition of relative entropy for $I_f$ we obtain the generalized Schr\"{o}dinger equation, which is unclear using the information geometry justification. Despite the difference between the present works and the entropic dynamics approach, it is encouraged to notice the common interests. In particular, the results in \cite{Ipek2019,Ipek2021} can be useful if we want to extend the present works to the scalar fields in curved spacetime.

The derivation of the Schr\"{o}dinger equation in Section \ref{sec:SE} starts from (\ref{extAction2}) which is inspired from its non-relativistic version initially proposed by Hall and Reginatto~\cite{Hall:2001,Hall:2002}. Ref.~\cite{Yang2023} gives a rigorous justification to the non-relativistic version of (\ref{extAction2}) using canonical transformation method. In Appendix \ref{appendix:canonical}, we extend the canonical transformation method to scalar fields and prove (\ref{extAction2}). Hall and Reginatto~\cite{Hall:2001,Hall:2002} only show the formulations in the non-relativistic setting. Even in the non-relativistic formulations, Hall and Reginatto assume an so-called exact uncertainty relation, while in our theory the  exact uncertainty relation is derived from the same least observability principle in a infinitesimal time step. 

\subsection{Limitations and Future Researches}
Assumption 1a makes minimal assumptions on the field fluctuations, but does not provide a more concrete physical model for the field fluctuations. The underlying physics for the field fluctuations is expected to be complex but crucial for a deeper understanding of quantum field theory. It is beyond the scope of this paper. The intention here is to minimize the assumptions that are needed to derive the  Schr\"{o}dinger equation for the wave functional, so that future research can just focus on justifying these assumptions. 

As shown in the appendix, the infinite dimension integration over the field variable $\phi(\mathbf{x})$ is approximated as a $N$ dimensional integral, then we take the limit $N\to\infty$. This essentially assumes a uniform Lebesgue measure. There is argument that probability integration measure is needed to ensure consistency between Fock representation and Sch\"{o}dinger representation~\cite{Corichi}. More rigorous mathematical treatment of infinite dimension integration is desirable. We also assume that the probability density $\rho[\phi]$ and its first order of functional derivative approach zero when $|\phi|\to\infty$. These assumptions are intuitive and give the correct results, but it is valuable to seek for stronger justifications.

The formulations presented in this paper is based on the flat Minkowski spacetime. We expect it is possible to extend the formulations to curved spacetimes and derive the Sch\"{o}dinger equation for curved spacetime. Furthermore, it would be interesting to investigate whether the least observability principle can be applied to non-scalar fields such as fermion matter fields whose equation of motion is the Dirac equation. 

\subsection{Conclusions}
The \added{extended least action principle, or }least observability principle, which is initially proposed to derive the non-relativistic quantum theory~\cite{Yang2023}, is applied here to the scalar field theory. We successfully obtain the Schr\"{o}dinger equation for the wave functional of the scalar field using the mathematical framework based on the principle. The Schr\"{o}dinger equation of the wave functional is the fundamental equation for the quantum scalar field theory in the Schr\"{o}dinger picture, and it is typically introduced as a postulate. Here we derive it from a first principle. The Schr\"{o}dinger equation enables one to calculate other standard results for the scalar fields, such as the solutions for the wave functional and the energy of the ground state and excited states\cite{Long, Jackiw}.

The least observability principle illustrates how classical field theory becomes quantum field theory from the information perspective. These are captured in the two assumptions stated in Section \ref{sec:LIP}. Assumption 2 points out that the Planck constant defines the discrete unit of action that a field configuration needs to exhibit in its dynamics in order to be observable. Classical field theory corresponds to a theory when such a lower limit of discrete action effort is approximated as zero. Assumption 1a demands new metrics to measure the additional observable information exhibited from field fluctuations, \added{which is then converted to additional action using Assumption 2.} These new information metrics are defined in terms of relative entropy to measure the information distances of different probability distributions caused by field fluctuations. To derive quantum theory, the extended least action principle seeks to minimize the total action from both classical field dynamics and additional field fluctuations. Nature appears to behave in a most economic fashion and exhibits as least observable information as possible. Furthermore, defining the information metrics $I_f$ using R\'{e}nyi divergence in the extended least action principle leads to a generalized Schr\"{o}dinger equation (\ref{SE2}) that depends on the order of R\'{e}nyi divergence. At this point it is inconceivable that one will find physical scenarios for which the generalized Schr\"{o}dinger equation for the wave functional with $\alpha \ne 1$ is applicable. However, the generalized Schr\"{o}dinger equation is legitimate from an information perspective. It confirms that the mathematical framework based on the extended least action principle can produce new results.

The works in Ref.~\cite{Yang2023} and this paper show that the extended least action principle can be applied to derive both non-relativistic quantum mechanics and relativistic quantum scalar field theory, demonstrating the versatility of the frameworks based on the principle. Extending the present work to scalar fields in curved spacetime is highly feasible. It is also reasonable to speculate the principle can be applied to obtain the quantum theory for non-scalar fields such as fermion matter fields, though it can be much more challenging since the structure of Lagrangian density for non-scalar fields is complicated. 

Lastly, the extended least action principle also brings in interesting implications on the interpretation aspects of quantum mechanics, including new insights on quantum entanglement, which will be reported separately. 


\added{
\begin{acknowledgements}
The author would like to thank the anonymous referees for their valuable comments, which help to strengthen the discussion of the extended least action principle and clarify the theory presented here compliant to relativistic requirements.
\end{acknowledgements}
}

\section*{Data Availability Statement}
The data that support the findings of this study are available within the article.





\onecolumngrid

\pagebreak

\appendix

\section{Canonical Transformation for Classical Scalar Field}
\label{appendix:canonical}
Suppose we choose a foliation of the Minkowksi spacetime
into a succession of fixed $t$ spacetime hypersurfaces $\Sigma_{t}$. The field configuration $\phi$ for $\Sigma_{t}$ can be understood as a vector with infinitely many components for each spatial point on the Cauchy hypersurface $\Sigma_t$ at time instance $t$ and, denoted as $\phi_{t,\mathbf{x}}=\phi(t,\mathbf{x}) = x$. Here, the meaning of $\phi(x)$ should be understood as the field component $\phi_{\mathbf{x}}$ at each spatial point of the hypersurfaces $\Sigma_{t}$ at time instance $t$. We want to transform from the pair of canonical variables $(\phi, \pi)$ into a generalized canonical variables $(\Phi, \Pi)$ and preserve the form of canonical equations. Denote the Lagrangian for both canonical coordinators as $L=\int_{\Sigma_t} \pi(x)\dot\phi(x) d^3\mathbf{x}-H(\phi, \pi)$ and $L'=\int_{\Sigma_t} \Pi(x)\dot\Phi(x) d^3\mathbf{x}-K(\Phi, \Pi)$, respectively, where $H$ is defined in (\ref{Hamiltonian}) and $K$ is the new form of Hamiltonian with the generalized canonical variables. We will omit the subscript ${\Sigma_t}$ in the integral. To ensure the form of canonical equations is preserved from the least action principle, one must have 
\begin{align}
    \delta \int^{t_B}_{t_A}dt L &= \delta\int^{t_B}_{t_A}dt (\int \pi(x)\dot\phi(x) d^3\mathbf{x}-H(\phi, \pi)) = 0\\
    \delta \int^{t_B}_{t_A}dt L' &= \delta\int^{t_B}_{t_A}dt (\int \Pi(x)\dot\Phi(x) d^3\mathbf{x}-K(\Phi, \Pi)) = 0.
\end{align}
One way to meet such conditions is that the Lagrangian in both integrals satisfy the following relation
\begin{equation}
    \label{extCan}
    \int \Pi(x)\dot\Phi(x) d^3\mathbf{x}-K(\Phi, \Pi) = \lambda (\int \pi(x)\dot\phi(x) d^3\mathbf{x}-H(\phi, \pi)) + \frac{dG}{dt},
\end{equation}
where $G$ is a generation functional, and $\lambda$ is a constant. When $\lambda \ne 1$, the transformation is called extended canonical transformations. Here we will choose $\lambda=-1$. Re-arranging (\ref{extCan}), we have
\begin{equation}
    \label{extCan2}
    \frac{dG}{dt} = \int (\Pi(x)\dot\Phi(x) + \pi(x)\dot\phi(x)) d^3\mathbf{x} - (K+H).
\end{equation}
Choose a generation functional $G=\int \Pi(x)\Phi(x) d^3\mathbf{x} + S(\phi, \Pi, t)$, that is, a type 2 generation functional analogous to the type 2 generation function in classical mechanics~\cite{Yang2023}. Its total time derivative is
\begin{equation}
    \label{type2}
    \frac{dG}{dt} = \int (\Pi(x)\dot\Phi(x) + \dot\Pi(x) \Phi(x)) d^3\mathbf{x} + \frac{\partial S}{\partial t} + \int (\frac{\delta S}{\delta\phi(x)})\dot\phi(x) d^3\mathbf{x} + \int (\frac{\delta S}{\delta\Pi(x)})\dot\Pi(x) d^3\mathbf{x}.
\end{equation}
The last two terms in (\ref{type2}) are obtained by applying the chain rule of functional derivative. Comparing (\ref{extCan2}) and (\ref{type2}) results in
\begin{align}
    \label{type12}
    \frac{\partial S}{\partial t} &= - (K+H), \\
    \pi(x) &= \frac{\delta S}{\delta\phi(x)}, \\
    \Phi(x) &= -\frac{\delta S}{\delta\Pi(x)}.
\end{align}
From (\ref{type12}), $K= - (\partial S/\partial t + H)$. Thus, $L'=\int \Pi(x)\dot\Phi(x) d^3\mathbf{x} + (\partial S/\partial t + H)$. We can choose a generation functional $S$ such that $\Phi$ does not explicitly depend on $t$ during motion. For instance, supposed $S(\phi, \Pi, t)=F(\phi, \Pi) + f(\phi, t)$, one has $\Phi=-\delta F(\phi, \Pi)/\delta\Pi(\mathbf{x})$, so that $\dot{\Phi}=0$ and $L' = \partial S/\partial t + H(\phi, \pi)$. Then the action integral in the generalized canonical coordinators becomes
\begin{equation}
    \label{extAction}
    A_c = \int^{t_B}_{t_A}dt L' = \int^{t_B}_{t_A}dt \{\frac{\partial S}{\partial t} + H(\phi, \pi)\}.
\end{equation}
where $H(\phi, \pi)$ is given in (\ref{Hamiltonian}). If one further imposes constraint on the generation functional $S$ such that the generalized Hamiltonian $K=0$, Eq. (\ref{type12}) becomes the field theory version of the Hamilton-Jacobi equation for the functional $S$, $\partial S/\partial t + H = 0$. It is a special solution for the least action principle based on $A_c$ when the generalized canonical field variables are $(\Phi, \Pi)$. 

Now consider that the field configuration $\phi$ is not definite but follows a probability distribution at any point of $\Sigma_t$. Alternatively, this can be understood as an ensemble of field configurations with probability density $\rho[\phi]$. In this case, the Lagrangian density is $\rho L'$, and the average value of the action integral for the ensemble of field configurations is,
\begin{equation}
    \label{extAction}
    S_c = \int \mathcal{D}\phi dt \{\rho(\phi) [\frac{\partial S}{\partial t} + H(\phi, \pi)]\},
\end{equation}
If we change the generalized canonical pair as $(\rho, S)$, applying the least action principle based on $S_c$ by variation of $S_c$ over $\rho$, one obtains, again, the field theory version of Hamilton-Jacobi equation for the functional $S$, $\partial S/\partial t + H = 0$.

\section{Proof of Eq. (\ref{varW})}
\label{appendix:varW}
Given the transition probability density (\ref{transP}), we want to calculate the normalization factor $Z$. There are an infinite number of spatial points in the hypersurface $\Sigma_t$. Rigorous mathematical treatment of infinite dimension integrals is challenged. We take a practical approach here and assume the fields are initially defined on a discrete lattice with $N$ number of vertices. Then, we take the limit of lattice distance approaching zero and $N\to\infty$. Equation (\ref{transP}) can be approximated as
\begin{equation}
    p[\omega] = \frac{1}{Z}e^{-\beta\sum_{i=1}^{N}\omega^2(x_i)\Delta x} = \frac{1}{Z}\prod_{i=1}^{N}e^{-\beta\omega^2(x_i)\Delta x},
\end{equation}
where $\Delta x =\Delta x^{(1)}\Delta x^{(2)}\Delta x^{(3)}$ is an infinitesimal small spatial volume, and $\beta=(\hbar\Delta t)^{-1}$. By the normalization condition,
\begin{equation}
    1=\int p[\omega]\mathcal{D}\omega = \frac{1}{Z}\int\prod_{i=1}^{N}e^{-\beta\omega^2(x_i)\Delta x}\prod_{j=1}^{N}d\omega(x_j)=\frac{1}{Z}\prod_{i=1}^{N}\int e^{-\beta\omega^2(x_i)\Delta x}d\omega(x_i)=\frac{1}{Z}(\frac{\beta\Delta x}{\pi})^{N/2}.
\end{equation}
Therefore, we have $Z=(\beta\Delta x/\pi)^{N/2}$. Next we evaluate $\langle \omega(x)\omega(x')\rangle$. Labeling the two spatial points $x=x_j$ and $x'=x_k$ in the lattice. If $j\ne k$,
\begin{align}
    \langle \omega(x)\omega(x')\rangle &= \int p[\omega]\omega(x_j)\omega(x_k)\mathcal{D}\omega \\
    & = \frac{1}{Z}\{\prod_{i\ne j, k}^{N}\int e^{-\beta\omega^2(x_i)\Delta x}d\omega(x_i)\}\{\int e^{-\beta\omega^2(x_j)\Delta x}\omega(x_j)d\omega(x_j)\}\{\int e^{-\beta\omega^2(x_k)\Delta x}\omega(x_k)d\omega(x_k)\}\\
    & = (\frac{\beta\Delta x}{\pi})^{-1}\int e^{-\beta\omega^2(x_j)\Delta x}\omega(x_j)d\omega(x_j)\int e^{-\beta\omega^2(x_k)\Delta x}\omega(x_k)d\omega(x_k).
\end{align}
But the two integrals are zero. Thus, $\langle \omega(x)\omega(x')\rangle = 0$. If $j=k$, similar calculation gives
\begin{equation}
    \langle \omega^2(x)\rangle = \sqrt{\frac{\beta\Delta x}{\pi}}\int e^{-\beta\omega^2(x_j)\Delta x}\omega^2(x_j)d\omega(x_j) = \frac{1}{2\beta\Delta x} = \frac{\hbar\Delta t}{2}\frac{1}{\Delta x}.
\end{equation}
Thus, we have
\begin{equation}
\label{deltaF}
    \langle \omega(x)\omega(x')\rangle= \left \{ \begin{array}{rcl} 0 & \mbox{for} & \mathbf{x}\ne \mathbf{x'} \\ \frac{\hbar\Delta t}{2}\frac{1}{\Delta x} & \mbox{for} & \mathbf{x}=\mathbf{x'}\end{array}\right.
\end{equation}
It is equivalent to rewrite (\ref{deltaF}) as $\langle \omega(x)\omega(x')\rangle = \frac{\hbar\Delta t}{2} \delta(\mathbf{x}-\mathbf{x'})$ since both expressions give the same identity
\begin{equation}
    \label{varW2}
    \int\langle \omega(x)\omega(x')\rangle d^3\mathbf{x'} = \frac{\hbar\Delta t}{2}.
\end{equation}

\section{Derivation of the Schr\"{o}dinger Equation}
\label{appendix:SE}
The key step in deriving the Schr\"{o}dinger equation is to prove (\ref{FisherInfo}) from (\ref{DLDivergence}). To do this, one first takes the functional derivative of $\rho[\phi+\omega]$ around $\phi$ up to the second order. Here we omit the time labeling for $\rho[\phi+\omega, t]$.
\begin{equation}
\label{Taylor}
    \rho[\phi+\omega] = \rho[\phi] + \int\frac{\delta\rho[\phi]}{\delta\phi(x)}\omega(x) d^3\mathbf{x} + \frac{1}{2}\int \frac{\delta^2\rho[\phi]}{\delta\phi(x)\delta\phi(x')} \omega(x)\omega(x')d^3\mathbf{x}d^3\mathbf{x'},
\end{equation}
The expansion is legitimate because (\ref{transP}) shows that the variance of fluctuation displacement $\omega$ is proportional to $\Delta t$. As $\Delta t \to 0$, only very small $w$ is significant. Then
\begin{align}
\label{Taylor2}
    ln\frac{\rho[\phi+\omega]}{\rho[\phi]} &= ln \{1 + \frac{1}{\rho}\int\frac{\delta\rho[\phi]}{\delta\phi(x)}\omega(x) d^3\mathbf{x} + \frac{1}{2\rho} \int \frac{\delta^2\rho[\phi]}{\delta\phi(x)\delta\phi(x')} \omega(x)\omega(x')d^3\mathbf{x}d^3\mathbf{x'}\} \\
    & = \frac{1}{\rho}\int\frac{\delta\rho[\phi]}{\delta\phi(x)}\omega(x) d^3\mathbf{x} + \frac{1}{2\rho} \int \frac{\delta^2\rho[\phi]}{\delta\phi(x)\delta\phi(x')}\omega(x)\omega(x')d^3\mathbf{x}d^3\mathbf{x'} -\frac{1}{2}\{\frac{1}{\rho}\int\frac{\delta\rho[\phi]}{\delta\phi(x)}\omega(x) d^3\mathbf{x}\}^2.
\end{align}
Substitute the above expansion into (\ref{DLDivergence}),
\begin{align*}
    \langle[D_{KL}(\rho[\phi, t_i] || \rho[\phi+\omega, t_i]]\rangle_{\omega} =& -\int\{\int\frac{\delta\rho[\phi]}{\delta\phi(x)}\langle\omega(x)\rangle d^3\mathbf{x} + \frac{1}{2} \int \frac{\delta^2\rho[\phi]}{\delta\phi(x)\delta\phi(x')}\langle\omega(x)\omega(x')\rangle d^3\mathbf{x}d^3\mathbf{x'} \\
    & -\frac{1}{2\rho}\langle\{\int\frac{\delta\rho[\phi]}{\delta\phi(x)}\omega(x) d^3\mathbf{x}\}^2\rangle\}\mathcal{D}\phi  \\
    =& -\int \{\frac{\hbar\Delta t}{4} \int \frac{\delta^2\rho[\phi]}{\delta\phi^2(x)} d^3\mathbf{x} -\frac{1}{2\rho}\langle\int\frac{\delta\rho[\phi]}{\delta\phi(x)}\omega(x) d^3\mathbf{x}\int\frac{\delta\rho[\phi]}{\delta\phi(x')}\omega(x') d^3\mathbf{x'}\rangle\}\mathcal{D}\phi\\
    =& - \int \{\frac{\hbar\Delta t}{4} \int \frac{\delta^2\rho[\phi]}{\delta\phi^2(x)} d^3\mathbf{x} -\frac{1}{2\rho}\int \frac{\delta\rho[\phi]}{\delta\phi(x)}\frac{\delta\rho[\phi]}{\delta\phi(x')}\langle\omega(x)\omega(x')\rangle d^3\mathbf{x}d^3\mathbf{x'}\}\mathcal{D}\phi\\
    = &- \frac{\hbar\Delta t}{4}\int \frac{\delta^2\rho[\phi]}{\delta\phi^2(x)} d^3\mathbf{x}\mathcal{D}\phi +\frac{\hbar\Delta t}{4}\int\frac{1}{\rho} (\frac{\delta\rho[\phi]}{\delta\phi(x)})^2d^3\mathbf{x}\mathcal{D}\phi
\end{align*}
In the above derivations, we have used the fact that $\langle \omega(x)\rangle=0$ and identity (\ref{varW}). Performing the integration in the first term by explicitly expanding the integration measure $\mathcal{D}\phi$ over all the spatial points $\mathbf{x'}$ in the hypersurface $\Sigma_t$,
\begin{align}
    \int \frac{\delta^2\rho[\phi]}{\delta\phi^2(x)} d^3\mathbf{x}\mathcal{D}\phi &= \int d^3\mathbf{x} \int \prod_{\mathbf{x'}\in\Sigma_t} d\phi_{\mathbf{x'}} \frac{\delta}{\delta\phi_{\mathbf{x}}}(\frac{\delta\rho}{\delta \phi_{\mathbf{x}}}) \\
    &= \int d^3\mathbf{x} \int \prod_{\mathbf{x'\ne x}} d\phi_{\mathbf{x'}} \int d\phi_{\mathbf{x}}\frac{\delta}{\delta\phi_{\mathbf{x}}}(\frac{\delta\rho}{\delta \phi_{\mathbf{x}}})\\
    \label{smoothRho}
    &=\int d^3\mathbf{x} \int \prod_{\mathbf{x'\ne x}} d\phi_{\mathbf{x'}} [(\frac{\delta\rho}{\delta \phi_{\mathbf{x}}})\vert_{\phi_{\mathbf{x}}=\infty} - (\frac{\delta\rho}{\delta \phi_{\mathbf{x}}})\vert_{\phi_{\mathbf{x}}=-\infty}].
\end{align}
Assuming $\rho$ is a smooth functional such that its first functional derivative approaches zero when $\phi_{\mathbf{x}}\to\pm\infty$, the above integral vanishes, and we obtain
\begin{equation}
    \langle[D_{KL}(\rho[\phi, t_i] || \rho[\phi+\omega, t_i])]\rangle_{\omega} = \frac{\hbar\Delta t}{4}\int\frac{1}{\rho[\phi, t_i]} (\frac{\delta\rho[\phi, t_i]}{\delta\phi(x)})^2d^3\mathbf{x}\mathcal{D}\phi.
\end{equation}
Substitute this into (\ref{DLDivergence}),
\begin{align}
    I_f &= \sum_{i=0}^{N-1}\langle[D_{KL}(\rho[\phi, t_i] || \rho[\phi+\omega, t_i])]\rangle_{\omega} \\
    &= \sum_{i=0}^{N-1} \frac{\hbar\Delta t}{4}\int\frac{1}{\rho[\phi, t_i]} (\frac{\delta\rho[\phi, t_i]}{\delta\phi(x)})^2d^3\mathbf{x}\mathcal{D}\phi\\
    \label{totalInfo4}
    &= \frac{\hbar}{4}\int\frac{1}{\rho[\phi, t]} (\frac{\delta\rho[\phi, t]}{\delta\phi(x)})^2d^3\mathbf{x}\mathcal{D}\phi dt,
\end{align}
which is Eq. (\ref{FisherInfo}). The next step is to derive (\ref{QHJ}). Variation of $I$ given in (\ref{totalDist}) with a small arbitrary change of  $\rho$, $\delta'\rho$, results in
\begin{align}
    \delta I =& \frac{2}{h}\int\{\frac{\partial S}{\partial t}\delta'\rho + \int\{ [\frac{1}{2}(\frac{\delta S}{\delta\phi(x)})^2 + V(\phi(x))]\delta'\rho + \frac{\hbar^2}{8}\delta'[\frac{1}{\rho} (\frac{\delta\rho}{\delta\phi(x)})^2]\}d^3\mathbf{x}\}\mathcal{D}\phi dt  \\
    \label{deltaI2}
    =& \frac{2}{h}\int\{\frac{\partial S}{\partial t}\delta'\rho + \int\{[\frac{1}{2}(\frac{\delta S}{\delta\phi(x)})^2 + V(\phi(x))]\delta'\rho - \frac{\hbar^2}{8}(\frac{1}{\rho}\frac{\delta\rho}{\delta\phi(x)})^2\delta'\rho + \frac{\hbar^2}{8}\frac{1}{\rho} \delta'(\frac{\delta\rho}{\delta\phi(x)})^2\}d^3\mathbf{x}\}\mathcal{D}\phi dt 
\end{align}
Note that the symbols $\delta'$ refers to variation over $\rho$ while $\delta$ refers to variation over $\phi$. Expanding the integration measure $\mathcal{D}\phi$ and performing the integration by part for the last term, we have
\begin{align}
    \int \frac{1}{\rho} \delta'(\frac{\delta\rho}{\delta\phi(x)})^2\}d^3\mathbf{x}\mathcal{D}\phi dt &= \int \frac{2}{\rho} (\frac{\delta\rho}{\delta\phi(x)})(\frac{\delta}{\delta\phi(x)}\delta'\rho)d^3\mathbf{x}\mathcal{D}\phi dt \\
    & = \int d^3\mathbf{x}\prod_{\mathbf{x'}\ne \mathbf{x}}d\phi_{\mathbf{x'}} dt\int d\phi_{\mathbf{x}}\frac{2}{\rho} (\frac{\delta\rho}{\delta\phi(x)})(\frac{\delta}{\delta\phi(x)}\delta'\rho) \\
    &= -\int d^3\mathbf{x}\prod_{\mathbf{x'}\ne \mathbf{x}}d\phi_{\mathbf{x'}} dt\int d\phi_{\mathbf{x}}(\frac{\delta}{\delta\phi(x)}[\frac{2}{\rho} (\frac{\delta\rho}{\delta\phi(x)})]\delta'\rho)\\
    \label{intByPart}
    &=-\int d^3\mathbf{x}\mathcal{D}dt\phi \{-[\frac{2}{\rho}\frac{\delta\rho}{\delta\phi(x)}]^2 +\frac{2}{\rho}\frac{\delta^2\rho}{\delta\phi^2(x)}\}\delta'\rho
\end{align}
Insert (\ref{intByPart}) back to (\ref{deltaI2}),
\begin{equation}
\label{deltaI3}
    \delta I = \frac{2}{h}\int\{\frac{\partial S}{\partial t} + \int\{ [\frac{1}{2}(\frac{\delta S}{\delta\phi(x)})^2 + V(\phi(x))] + \frac{\hbar^2}{8}[(\frac{1}{\rho}\frac{\delta\rho}{\delta\phi(x)})^2 - \frac{2}{\rho}\frac{\delta^2\rho}{\delta\phi^2(x)}]\} d^3\mathbf{x}\}\delta'\rho\mathcal{D}\phi dt.
\end{equation}
Taking $\delta I = 0$ for arbitrary $\delta'\rho$, we must have
\begin{equation}
\label{QHJ2}
    \frac{\partial S}{\partial t} + \int\{ [\frac{1}{2}(\frac{\delta S}{\delta\phi(x)})^2 + V(\phi(x))] + \frac{\hbar^2}{8}[(\frac{1}{\rho}\frac{\delta\rho}{\delta\phi(x)})^2 - \frac{2}{\rho}\frac{\delta^2\rho}{\delta\phi^2(x)}]\} d^3\mathbf{x}=0.
\end{equation}
Defining $R[\phi,t]=\sqrt{\rho[\phi, t]}$, one can verify that 
\begin{equation}
    (\frac{1}{\rho}\frac{\delta\rho}{\delta\phi(x)})^2 - \frac{2}{\rho}\frac{\delta^2\rho}{\delta\phi^2(x)} = - \frac{4}{R}\frac{\delta^2 R}{\delta\phi^2(x)}.
\end{equation}
Substituting it back to (\ref{QHJ2}) gives the desired result in (\ref{QHJ}). Now defining $\Psi[\phi, t]=\sqrt{\rho[\phi, t]}e^{iS/\hbar}$, and substituting (\ref{QHJ2}) and the continuity equation (\ref{contEq}), we have
\begin{align}
    \frac{i\hbar}{\Psi}\frac{\partial \Psi}{\partial t} &= \frac{i\hbar}{2\rho}\frac{\partial \rho}{\partial t} - \frac{\partial S}{\partial t}\\
    \label{SE11}
    &= -\frac{i\hbar}{2\rho}\int \frac{\delta}{\delta\phi(x)}(\rho\frac{\delta S}{\delta\phi(x)}) d^3\mathbf{x} + \int\{ [\frac{1}{2}(\frac{\delta S}{\delta\phi(x)})^2 + V(\phi(x))] + \frac{\hbar^2}{8}[(\frac{1}{\rho}\frac{\delta\rho}{\delta\phi(x)})^2 - \frac{2}{\rho}\frac{\delta^2\rho}{\delta\phi^2(x)}]\} d^3\mathbf{x}.
\end{align}
On the other hand, computing the second order of functional derivative of $\Psi$ gives
\begin{align}
    \frac{\delta\Psi}{\delta\phi(x)} &= \frac{1}{2\rho}\frac{\delta\rho}{\delta\phi(x)}\Psi + \frac{i}{\hbar}\frac{\delta S}{\delta\phi(x)}\Psi \\
    \frac{\delta^2\Psi}{\delta\phi^2(x)} &= \frac{i}{\hbar}\frac{1}{\rho}\frac{\delta}{\delta\phi(x)}(\rho\frac{\delta S}{\delta\phi(x)})\Psi -\frac{1}{4}[(\frac{1}{\rho}\frac{\delta\rho}{\delta\phi(x)})^2 - \frac{2}{\rho}\frac{\delta^2\rho}{\delta\phi^2(x)}]\Psi -\frac{1}{\hbar^2}(\frac{\delta S}{\delta\phi(x)})^2\Psi\\
    \label{SE12}
    -\frac{\hbar^2}{2}\frac{\delta^2\Psi}{\delta\phi^2(x)} &= -\frac{i\hbar}{2\rho}\frac{\delta}{\delta\phi(x)}(\rho\frac{\delta S}{\delta\phi(x)})\Psi +\frac{\hbar^2}{8}[(\frac{1}{\rho}\frac{\delta\rho}{\delta\phi(x)})^2 - \frac{2}{\rho}\frac{\delta^2\rho}{\delta\phi^2(x)}]\Psi +\frac{1}{2}(\frac{\delta S}{\delta\phi(x)})^2\Psi.
\end{align}
Comparing (\ref{SE11}) and (\ref{SE12}), one can recognize the Schr\"{o}dinger equation for the wave functional $\Psi$,
\begin{equation}
    \frac{i\hbar}{\Psi}\frac{\partial \Psi}{\partial t} = \int [-\frac{\hbar^2}{2}\frac{\delta^2}{\delta\phi^2(x)} + V(\phi(x))]d^3\mathbf{x}.
\end{equation}

\section{R\'{e}nyi Divergence and the Generalized Schr\"{o}dinger Equation}
\label{appendix:RE}
Based on the definition of $I_f^{\alpha}$ in (\ref{RDivergence}), and starting from (\ref{Taylor}), we have
\begin{align*}
    \int\mathcal{D}\phi\frac{\rho^{\alpha}[\phi]}{\rho^{\alpha -1}[\phi+\omega]} =& \int\mathcal{D}\phi\rho(1+\frac{1}{\rho}\int\frac{\delta\rho[\phi]}{\delta\phi(x)}\omega(x) d^3\mathbf{x} + \frac{1}{2\rho}\int \frac{\delta^2\rho[\phi]}{\delta\phi(x)\delta\phi(x')} \omega(x)\omega(x')d^3\mathbf{x}d^3\mathbf{x'})^{1-\alpha} \\
    =& \int\mathcal{D}\phi\rho\{1+(1-\alpha)(\frac{1}{\rho}\int\frac{\delta\rho[\phi]}{\delta\phi(x)}\omega(x) d^3\mathbf{x} + \frac{1}{2\rho}\int \frac{\delta^2\rho[\phi]}{\delta\phi(x)\delta\phi(x')} \omega(x)\omega(x')d^3\mathbf{x}d^3\mathbf{x'})\\
    & +\frac{1}{2}\alpha(\alpha-1)(\frac{1}{\rho}\int\frac{\delta\rho[\phi]}{\delta\phi(x)}\omega(x) d^3\mathbf{x} + \frac{1}{2\rho}\int \frac{\delta^2\rho[\phi]}{\delta\phi(x)\delta\phi(x')} \omega(x)\omega(x')d^3\mathbf{x}d^3\mathbf{x'})^2\} \\
    =& 1 + (1-\alpha)\int(\int\frac{\delta\rho[\phi]}{\delta\phi(x)}\omega(x) d^3\mathbf{x} + \frac{1}{2}\int \frac{\delta^2\rho[\phi]}{\delta\phi(x)\delta\phi(x')} \omega(x)\omega(x')d^3\mathbf{x}d^3\mathbf{x'})\mathcal{D}\phi \\
    & + \frac{1}{2}\alpha(\alpha-1)\int\frac{1}{\rho}(\int\frac{\delta\rho[\phi]}{\delta\phi(x)}\omega(x) d^3\mathbf{x})^2\mathcal{D}\phi.
\end{align*}
In the last step we have applied the normalization condition $\int \rho \mathcal{D}\phi = 1$. Similar to the derivation of (\ref{smoothRho}), by assuming $\rho$ is a smooth functional such that its values and first functional derivative approaches zero when $\phi_{\mathbf{x}}\to\pm\infty$, the second term of the above equation vanishes after performing the integration over $\mathcal{D}\phi$. Then, we have
\begin{align*}
   ln\{ \int\frac{\rho^{\alpha}[\phi]}{\rho^{\alpha -1}[\phi+\omega]}\mathcal{D}\phi \}  &= ln \{ 1 + \frac{1}{2}\alpha(\alpha-1)\int \frac{1}{\rho}(\int\frac{\delta\rho[\phi]}{\delta\phi(x)}\omega(x) d^3\mathbf{x})^2 \mathcal{D}\phi\}\\
   &= \frac{1}{2}\alpha(\alpha-1)\int \frac{1}{\rho}(\int\frac{\delta\rho[\phi]}{\delta\phi(x)}\omega(x) d^3\mathbf{x})^2 \mathcal{D}\phi.
\end{align*}
Thus, $I_f^{\alpha}$ is simplified as
\begin{align}
\label{RDivergence2}
    I_f^{\alpha} 
    &=\sum_{i=0}^{N-1}\langle\frac{1}{\alpha-1}ln\{ \int\frac{\rho^{\alpha}[\phi, t_i]}{\rho^{\alpha -1}[\phi+\omega, t_i]}\mathcal{D}\phi \}\rangle_{\omega} \\
    & = \sum_{i=0}^{N-1}\langle\frac{\alpha}{2} \frac{1}{\rho}(\int\frac{\delta\rho[\phi]}{\delta\phi(x)}\omega(x) d^3\mathbf{x})^2 \mathcal{D}\phi\rangle_{\omega} \\
    & = \sum_{i=0}^{N-1}\frac{\alpha}{2} \int \frac{1}{\rho} (\int\frac{\delta\rho[\phi]}{\delta\phi(x)}\frac{\delta\rho[\phi]}{\delta\phi(x')}\langle\omega(x)\omega(x')\rangle_{\omega} d^3\mathbf{x}d^3\mathbf{x'})\mathcal{D}\phi\\
\label{I_f4}
    &= \sum_{i=0}^{N-1}\frac{\alpha\hbar}{4}\Delta t\int \frac{1}{\rho}(\frac{\delta\rho[\phi]}{\delta\phi(x)})^2 d^3\mathbf{x}\mathcal{D}\phi= \frac{\alpha\hbar}{4}\int  \frac{1}{\rho}(\frac{\delta\rho[\phi]}{\delta\phi(x)})^2 d^3\mathbf{x}\mathcal{D}\phi dt.
\end{align}
Compared to (\ref{totalInfo4}), the only difference from $I_f$ is that there is an additional coefficient $\alpha$, i.e., $I_f^{\alpha} = \alpha I_f$. The total observability is 
\begin{equation}
    \label{totalDist2} 
    I = \frac{2}{h}\int\rho\{\frac{\partial S}{\partial t} + \int [\frac{1}{2}(\frac{\delta S}{\delta\phi(x)})^2 + V(\phi(x))  + \frac{\alpha\hbar^2}{8}(\frac{1}{\rho}\frac{\delta\rho}{\delta\phi(x)})^2 ]d^3\mathbf{x}\}\mathcal{D}\phi dt.
\end{equation}
Fixed end point variation of $I$ with respect to $S$ gives the same continuity equation (\ref{contEq}). Variation with respect to $\rho$ by following the same calculations from (\ref{deltaI2}) to (\ref{deltaI3}) in Section \ref{appendix:SE} leads to (\ref{RHJ}). Defined $\hbar_{\alpha}=\sqrt{\alpha}\hbar$, equation (\ref{RHJ}) can be rewritten as 
\begin{equation}
\label{RHJ'}
\begin{split}
    \frac{\partial S}{\partial t} =- \int\{\frac{1}{2}(\frac{\delta S}{\delta\phi(x)})^2 + V(\phi(x)) - \frac{\hbar_{\alpha}^2}{2R}\frac{\delta^2 R}{\delta\phi^2(x)} \}d^3\mathbf{x},
    \end{split}
\end{equation}
Equation (\ref{RHJ'}) is in the same form as (\ref{QHJ}) except replacing $\hbar$ with $\hbar_{\alpha}$. Notice that the continuity equation does not contain the Planck constant. Since the Schr\"{o}dinger equation simply combines the continuity equation and (\ref{RHJ'}) by defining a complex functional $\Psi_{\alpha}[\phi, t]=R[\phi, t]e^{iS/\hbar_{\alpha}}$, performing the similar calculations in (\ref{SE11}) and (\ref{SE12}), we obtain the same form of Schr\"{o}dinger equation but with $\hbar$ replaced by $\hbar_{\alpha}$,
\begin{equation}
    \label{SE'}
    i\hbar_{\alpha}\frac{\partial\Psi_{\alpha}[\phi, t]}{\partial t} = \{\int[-\frac{\hbar_{\alpha}^2}{2}\frac{\delta^2}{\delta\phi^2(x)} + V(\phi(x))]d^3\mathbf{x}\}\Psi_{\alpha}[\phi, t].
\end{equation}
Replacing back $\hbar_{\alpha}=\sqrt{\alpha}\hbar$ in (\ref{SE'}) gives (\ref{SE2}).

\end{document}